\documentclass[conference,a4paper]{IEEEtran}
\usepackage{xcolor}
\usepackage{balance}

\ifCLASSINFOpdf
  \usepackage[pdftex]{graphicx}
\else
  \usepackage[dvips]{graphicx}
\fi
\ifCLASSOPTIONcompsoc
 \usepackage[caption=false,font=normalsize,labelfont=sf,textfont=sf]{subfig}
\else
 \usepackage[caption=false,font=footnotesize]{subfig}
\fi
\usepackage{hft-thesis}

\AddToHook{shipout/foreground}{
  \put(0.5\paperwidth, -0.8cm){ 
    \makebox[0pt][c]{
      \begin{minipage}{\textwidth}
        \centering \scriptsize 
          This article has been accepted for publication in 20th European Conference on Antennas and Propagation (EuCAP). This is the author's version which has not been fully edited and content may change prior to final publication.
      \end{minipage}
    }
  }
}

\AddToHookNext{shipout/foreground}{
  \put(0.5\paperwidth, -\paperheight + 0.7cm){ 
    \makebox[0pt][c]{
      \begin{minipage}{\textwidth}
        \centering \scriptsize
          Copyright~\copyright~2026 IEEE. Personal use of this material is permitted. Permission from IEEE must be obtained for all other uses, in any current or future media, including\\reprinting/republishing this material for advertising or promotional purposes, creating new collective works, for resale or redistribution to servers or lists, or reuse of any copyrighted component of this work in other works by sending a request to pubs-permissions@ieee.org.
      \end{minipage}
    }
  }
}

\begin{document}
%
\title{Phase-Corrected Near-Field Microwave Imaging \\ via Inverse Source Reconstruction \\ with Modulated Signals}

\author{\IEEEauthorblockN{
Quanfeng Wang, Alexander H. Paulus, and Thomas F. Eibert   
}                                     
\IEEEauthorblockA{
  Department of Electrical Engineering, School of Computation, Information and Technology,\\
  Technical University of Munich, Munich, Germany, quanfeng.wang@tum.de}
}



\maketitle

\begin{abstract}
An inverse source reconstruction (ISR) based \mbox{3-D} near-field (NF) passive radar microwave imaging method utilizing modulated signals is presented. The modulated signals from a non-cooperative transmitter are scattered by the targets of interest and captured by a fixed reference antenna together with an NF scanning probe at different positions. By normalizing with the reference signals, spatial coherence of the NF observations is obtained, and a single-frequency inverse source solver is subsequently utilized for ISR and image generation. A corresponding phase correction method is proposed for the coherent superposition of multi-frequency images and verified through simulations. In addition, it is shown that for realistic narrowband signals, an incoherent imaging approach is sufficient. The presented technical scheme is validated using a planar scanning system in a typical office room, where software-defined radios are employed for the transmitting and receiving of narrowband orthogonal frequency-division multiplexing signals at Wi-Fi operating frequencies. With the aid of background subtraction and reference signals, images of a mannequin placed in the office room are successfully obtained.
\end{abstract}

\vskip0.5\baselineskip
\begin{IEEEkeywords}
 Microwave imaging, modulated signals, OFDM, passive radar, phase correction.
\end{IEEEkeywords}

%

\section{Introduction}
Microwave imaging has gained increasing popularity in recent years, particularly also the concept of passive radar which was first introduced in 1935~\cite{griffiths2010klein,kuschel2019tutorial}. In contrast to most active microwave imaging systems, which require a dedicated transmitter (Tx) and receiver (Rx) or a cooperative multiple-input multiple-output synthetic aperture radar (SAR) configuration, passive radar utilizes the ubiquitous radiation from existing radio-frequency technologies.

The use of a reference antenna is common in both near-field (NF) and far-field (FF) passive radar imaging for purposes such as synchronization, direct signal interference, and clutter cancellation~\cite{colone2023passive}. In~\cite{huang2022passive}, a static reference antenna steered toward the satellites is used to obtain signals from a global navigation satellite system. Cross correlation is then applied to achieve signal synchronization of the scattered echo from the FF surveyed maritime area. A similar approach is applied in~\cite{samczynski20225g} to 5G network-based FF passive radar, where the cross-ambiguity function for target detection relies on a reference channel. For NF Wi-Fi imaging in~\cite{holl2017holography}, the holographic data are recorded by a scanning probe and normalized using a reference antenna. 3-D images are then obtained through numerical backpropagation and incoherent averaging over a $\SI{70}{\mega\hertz}$ bandwidth centered around the carrier frequency at $\SI{5}{\giga \hertz}$ in accordance with IEEE 802.11ac standard protocols. 

Although efforts have been made toward reference-free or even incoherent measurement approaches in passive radar imaging, their performance remains limited and less robust. For instance, Wision~\cite{huang2014feasibility} performs reference-free imaging based on the coherent combination of scattered signals at specific azimuth and elevation angles. However, this method requires an antenna array to ensure spatial coherence and can only be applied to large objects due to its poor resolution. \mbox{Wi-Vi}~\cite{adib2013see}, which consists of two Tx antennas and one Rx antenna, is designed for through-wall imaging. It needs sophisticated control and is limited to the imaging of moving targets. In~\cite{vakalis2019imaging}, a passive interferometric imaging method is presented that enables imaging with incoherent measurements, but it relies on multiple Wi-Fi transmitters emitting signals that are both spatially and temporally incoherent. 

An inverse source reconstruction (ISR) based NF passive radar imaging method has been presented in~\cite{wang2024TAP,wang2024microwavea}. A sufficient number of NF samples are recorded through planar measurements, and a single-frequency inverse equivalent source solver~\cite{eibert2015electromagnetic} is then applied to reconstruct the equivalent sources in the form of plane wave spectra (PWS). Single-frequency images are subsequently generated from the obtained PWS for each individual frequency and coherently superimposed using a corresponding phase correction method. Although proven effective, this technique has so far been verified only under continuous wave NF illumination. In this work, the method is extended to modulated signals to meet the requirements of realistic passive radar applications. A fixed reference antenna continuously receives reference signals during the repeated planar scanning, and the acquired NF samples are normalized by the corresponding reference signals to achieve spatial coherence. Consequently, the ISR process can be applied for the imaging, but, due to the normalization, the phase correction procedure must also be adapted for coherent multi-frequency imaging.

In Section~\ref{sec:img_gen}, the imaging algorithm and the corresponding phase correction method are introduced. In Section~\ref{sec:simu}, numerical results and further discussion of the phase correction method for both narrow- and wideband signals are provided. Measurement results are found in Section~\ref{sec:meas}, and some conclusions are drawn in Section~\ref{sec:conclusion}.

\section{Imaging Algorithm and Phase Correction}
\label{sec:img_gen}
\begin{figure}[t]
	\centering
	\includegraphics[scale=0.6]{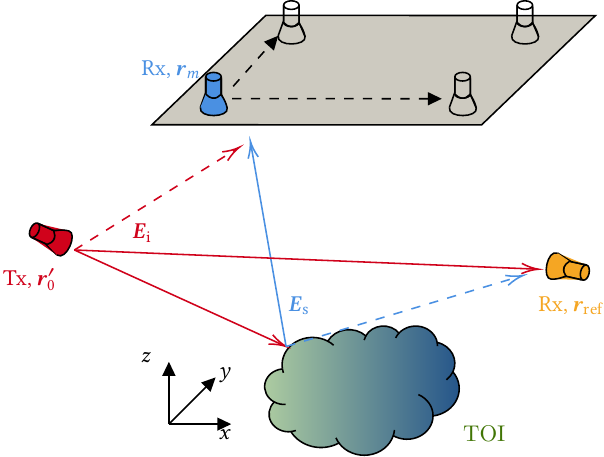}
	\caption{Imaging configuration of the TOI with a single Tx antenna and a fixed reference antenna. The scattered fields and incident fields are recorded by a planar NF observation setup.}
	\label{fig:configuration}
\end{figure}%

Considering the imaging configuration shown in Fig.~\ref{fig:configuration}, a fixed Tx antenna illuminates the targets of interest (TOI), while an Rx antenna continuously scans the electric field at different locations $\veg{r}_m$ on a plane. Meanwhile, a fixed reference antenna located at $\veg{r}_{\mathrm{ref}}$ receives the reference signals, which may consist of both the incident and scattered fields. In practical measurements inphase and quadrature (I/Q) samples over time are collected, and Fourier transforms are performed to obtain the corresponding frequency domain harmonic samples $\veg{E}(\veg{r}_m, f)$ and $\veg{E}_{\mathrm{ref}}(\veg{r}_m, f)$, where a harmonic time-dependence $\e^ {\jm 2\uppi ft}$ is assumed and suppressed. Moreover, the frequency dependence of fields and currents is not written in the following. The ISR and the subsequent image generation are then carried out for each individual frequency. The modulated incident field can be characterized by an equivalent volume current density $\veg{J}_{ \rm i}\left(\veg{r'}\right)$ in the form of
\begin{align}
  \veg{E}_{\rm i}\left({\veg{r}_m},B_{m}\left(f\right)\veg{J}_{\rm i}\right)= &\iiint_{V_{\rm i}}  \vecop{G}\left(\veg{r}_m,\veg{r'}\right)\cdot \left[B_{m}(f)\veg{J}_{ \rm i}\left(\veg{r'}\right)\right]\,\dd v' \notag \\ 
                    =&  B_{m}(f)\veg{E}_{\rm i}\left({\veg{r}_m},\veg{J}_{\rm i}\right) \,, \label{eq:first}
\end{align}
where $\vecop{G}$ denotes the dyadic free-space Green's function and $B_{m}(f)$ is an unknown frequency-dependent complex coefficient that characterizes the magnitude and phase variations due to the modulation as well as the missing Tx-Rx synchronization at frequency $f$, which may also vary across different measurement positions. With the assumption of the Born approximation~\cite{newton2002scattering}, the current density $\veg{J}_{\rm s}\left(\veg{r'}\right)$ induced in the TOI is related to the incident source density $\veg{J}_{\rm i}\left(\veg{r'}\right)$ and the scattering dyad $\overset{\leftrightarrow }{\bm \sigma }$ 
\begin{equation}
  \veg{J}_{\rm s}\left(\veg{r'}\right)\approx \overset{\leftrightarrow }{\bm \sigma }\left({\veg{r'}}\right) \cdot B_{m}(f) \veg{E}_{\rm i}\left({\veg{r'}},\veg{J}_{\rm i}\right)\,.\label{eq:ohm}
\end{equation}
Thus, with the linearization, the NF samples are expressed by the sum of the incident and scattered fields
\begin{align}
  \veg{E}(\veg{r}_m, B_{m}\left(f\right)\veg{J}_{\rm i})=&B_{m}(f) \left[\veg{E}_{\rm i}\left({\veg{r}_m},\veg{J}_{\rm i}\right)+\veg{E}_{\rm s}\left({\veg{r}_m},\overset{\leftrightarrow }{\bm \sigma }\left({\veg{r'}}\right) \cdot \veg{E}_{\rm i}\left({\veg{r'}},\veg{J}_{\rm i}\right)\right)\right]\notag \\
  =&B_{m}(f) \veg{E'_{\rm tot}}(\veg{r}_m,\veg{J}_{\rm i})\,,
\end{align}
where $\veg{E'_{\rm tot}}(\veg{r}_m, \veg{J}_{\rm i})$ denotes the total electric field at $\veg{r}_m$ generated by the sources $\veg{J}_{\rm i}$ under the Born approximation without any modulation. Since the same formulation can be obtained for the reference signal, the modulation coefficient in the NF measurements can be effectively suppressed by normalizing with the reference signal, i.e.,
\begin{equation}
  \bar{\veg{E}}\left(\veg{r}_m\right)=\frac{\veg{E}(\veg{r}_m, B_{m}\left(f\right)\veg{J}_{\rm i})}{{E_{p}}(\veg{r}_{\rm{ref}}, B_{m}\left(f\right)\veg{J}_{\rm i})}= \frac{\veg{E'}_{\rm tot}(\veg{r}_{m}, \veg{J}_{\rm i})}{{E'_{\mathrm{tot}, \,p}}(\veg{r}_{\rm ref},  \veg{J}_{\rm i})}\,,\label{eq:norm}
\end{equation}
where a certain Cartesian vector component $p \in\{x,y,z\}$ is selected for the normalization in order to preserve the polarization of the NF samples. With the normalization, spatial coherence is obtained, no matter whether modulation is present or not, and the ISR can then be applied.

By expanding the Green's function into propagating plane waves and representing the equivalent sources of the Tx and the TOI by its spectra $\tilde{\veg{J}_{\rm i}}(\veg{k},\veg{r}'_{\rm i})$ and $\tilde{\veg{J}}_{\rm s}(\veg{k},\veg{r}'_{\rm s})$, which are most efficiently derived from corresponding equivalent surface source representations on appropriately chosen Huygens' surfaces, a linear system of equations can be formulated for each discrete frequency based on~\cite{wang2024TAP,eibert2015electromagnetic,Chew2001}
\begin{align}
   \bar{\veg{E}}\left(\veg{r}_m\right)=\frac{-\jm}{4\uppi}  \oiint &\left[\;\; \sum_{\veg{r'_{\rm{i}}}} T_{L}\left(\veg{k},\veg{r}_{m}-\veg{r'_{\rm{i}}}\right)\,{\tilde{\veg{J}_{\rm i}}(\veg{k},\veg{r}'_{\rm i})}\notag\right.\\ \phantom{\left[\sum_{\veg{r}'_{\rm{s}}}\right]} &\left. +\sum_{\veg{r}'_{\rm{s}}}T_{L}\left(\veg{k},\veg{r}_{m}-\veg{r}'_{\rm{s}}\right)\,{\tilde{\veg{J}}_{\rm s}(\veg{k},\veg{r}'_{\rm s})}\right]\;\dd^2 \hat{k}     \,,\label{eq:FMM}
\end{align}
where $\veg{k}=k\hat{ k}$ denotes the wave vector and $T_L$ of order $L$ is the well-known fast multipole translation operator~\cite{Chew2001}. With the obtained PWS of the TOI, the polarimetric single-frequency images can be computed by hierarchical disaggregation~\cite{wang2024TAP,schnattinger2012solution},
\begin{equation}
  {\mathring{\veg{J}}_{\rm{s}}\left(k_{f},\veg{r}'\right)}=\oiint \mathcal{F}\left(\hat{k}\cdot\hat{k}^{\mathrm{(c)}}\right) {\tilde{\veg{J}}_{\rm{s}}(\veg{k}_f,\veg{r_{\rm{s}}})} \e^{-\jm \veg{k}_f \cdot \veg{r}'} \, \dd^2 \hat{k}\,,\label{eq:imageGen}
\end{equation}
where $\mathcal{F}\left(\hat{k}\cdot\hat{k}^{\mathrm{(c)}}\right)$ is an angular spectral windowing function balancing the resolution and artifacts~\cite{wang2024TAP,eibert2023inverse}.

For coherent superposition across all frequencies, the phase and magnitude correction terms $\psi_{\mathrm{s}}\left(k_f,\veg{r}'\right)$ and $\mathcal{M}_{\mathrm{ s}}\left(k_f, \veg{r}'\right)$ introduced in~\cite{wang2024TAP} are required. An explanation for $\psi_{\mathrm{s}}\left(k_f,\veg{r}'\right)$ under the Born approximation is that it compensates for the influence of $\veg{E}_{\rm i}$ in \eqref{eq:ohm}, which is achieved either by assuming a dipole source~\cite{wang2024TAP} or by adopting a simpler yet similar effective isotropic point source assumption~\cite{wang2025automatic}.
In addition, the frequency-dependent phase variation in the denominator of~\eqref{eq:norm} is still present after the image generation due to the linear relationship in \eqref{eq:FMM} and \eqref{eq:imageGen}, and must, therefore, also be compensated. Taking again the isotropic point source assumption and requiring the incident field received by the reference antenna to be much stronger than the scattered field, the second phase correction term in addition to $\psi_{\mathrm{s}}\left(k_f,\veg{r}'\right)$ can be denoted as
\begin{equation}
  \psi_{\mathrm{ref}}\left(k_f\right)=\e^{-\jm k_{f} \left(\left|\veg{r}_{\rm ref}-\veg{r}'_{0}\right|\right)}\,.
\end{equation}
In this work, the same phase corrections are utilized for all polarimetric images, however, they can also be applied separately to different polarizations if desired~\cite{wang2024TAP}. Consequently, the final multi-frequency image generation is expressed as
\begin{align}
   \veg{J}(\veg{r}')=\sum_{f}&\psi_{\mathrm{s}}\left(k_f,\veg{r}'\right)\psi_{\mathrm{ref}}\left(k_f\right){\mathcal{M}_{\mathrm{ s}}(k_f,\veg{r}')} {\mathring{\veg{J}}_{\mathrm{s}}\left(k_{f},\veg{r}'\right)}\,. \label{eq:sumF}
\end{align}

\section{Numerical Results}
\label{sec:simu}
\begin{figure}[t]
	\centering
	\subfloat[]{\includegraphics[scale=0.28]{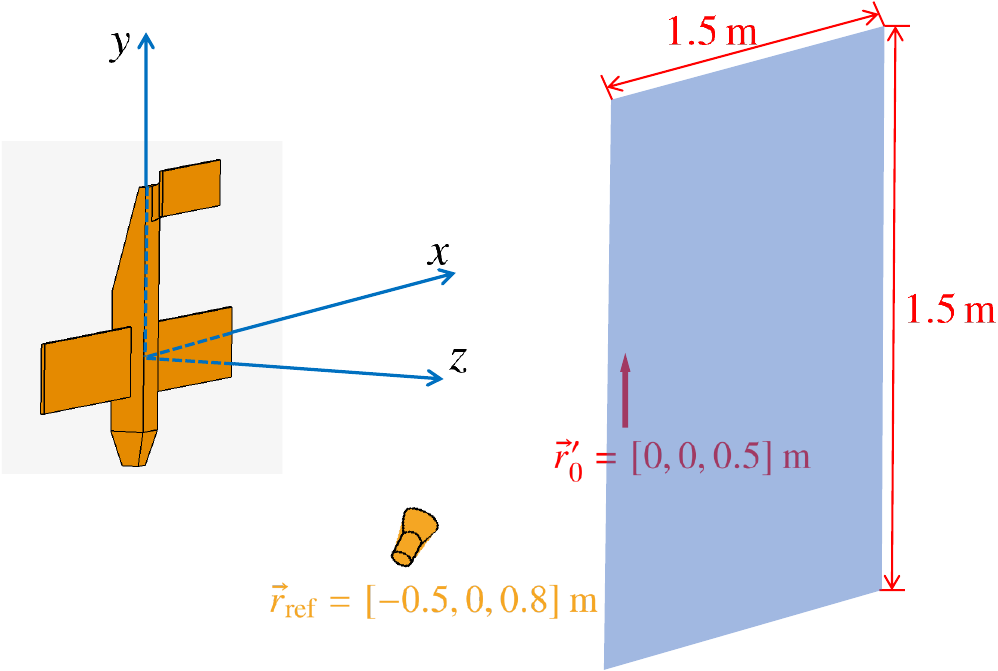}}%
	\hfill
	\subfloat[]{\includegraphics[scale=0.32]{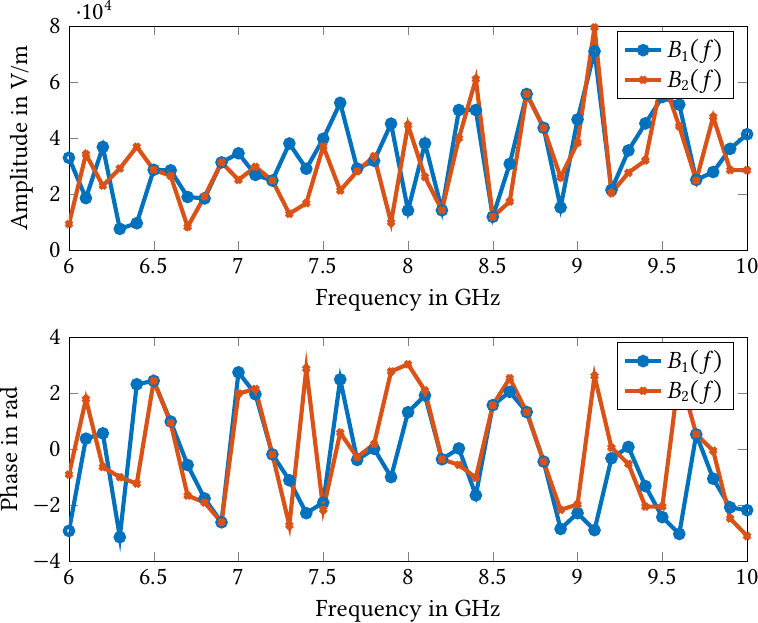}}%
	\caption{(a) Illustration of the simulation configuration with a PEC aircraft located around the origin. A Hertzian dipole located at $\veg{r}'_{0}$ is used as the illumination source and a reference antenna is placed at $\veg{r}_{\rm ref}$. (b)~ Reference signals exhibit different amplitude and phase variations at two different locations.}
	\label{fig:Simu_setup}
\end{figure}
Full-wave simulations were conducted using the commercial software FEKO~\cite{EMSS2024}. The configuration consisting of a perfectly electrically conducting (PEC) aircraft is shown in Fig.\,\ref{fig:Simu_setup}(a), where the Tx is modeled as a $y$-polarized Hertzian dipole located at $\veg{r}'_{0}=[0, 0, 0.5]\,$m, and the $y$-polarized electric field used as the reference signal is extracted at $\veg{r}_{\rm ref}=[-0.5, 0, 0.8]\,$m. The simulations were performed with a frequency domain solver, and modulation of the fields was realized by multiplying the Tx excitation with a random magnitude and phase variation characterized by $B_m(f)$. In practice, a typical Tx transmits time-varying data packets, and different I/Q samples are captured at each probing position $\veg{r}_{m}$ during the NF scanning. Accordingly, in our simulations, a different $B_m(f)$ was assigned to each $\veg{r}_{m}$, and of course also to the corresponding reference signal. In total, $10\,000$ uniformly distributed NF samples were collected on a plane at $z=\SI{1}{\meter}$, spanning from $[x, y]=[-0.75, -0.75]\,$m to $[x, y]=[0.75, 0.75]\,$m. Therefore, $10\,000$ unique variations of $B_m(f)$ were applied in the simulation. Two representative examples are shown in Fig.~\ref{fig:Simu_setup}(b). 

\begin{figure}[t]
  \centering
  \subfloat[]{\includegraphics[scale=0.5]{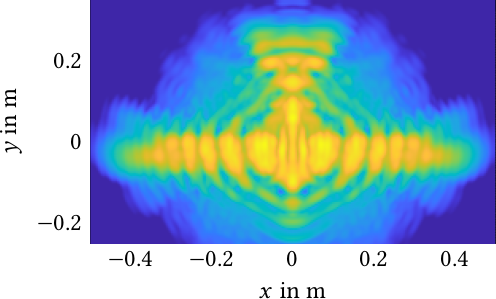}}%
  \hfill
  \subfloat[]{\includegraphics[scale=0.5]{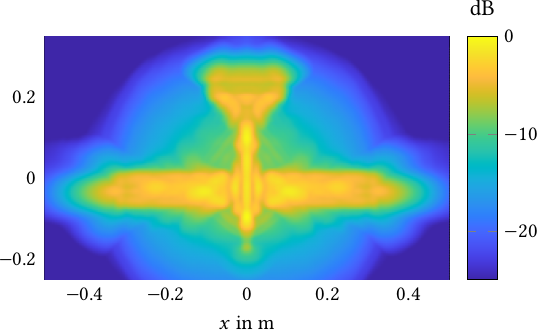}}%
  \\ 
  \vspace{-5mm}
  \subfloat[]{\includegraphics[scale=0.65]{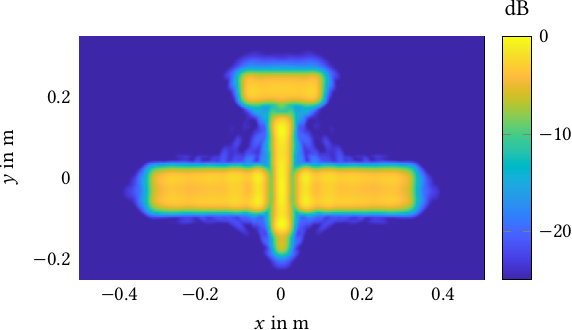}}%
  \caption{Imaging results of the PEC aircraft as 2D MIP in the $xy$-plane, (a)~single-frequency image at $\SI{8}{\giga\hertz}$, (b)~incoherent multi-frequency image, and (c)~coherent multi-frequency image.}
  \label{fig:plane_res}
\end{figure}

For the first simulation, a relatively wide frequency range from $\SI{6}{\giga\hertz}$ to $\SI{10}{\giga\hertz}$ with a constant step size of $\SI{100}{\mega\hertz}$ was considered. The imaging results for a single-frequency image at $\SI{8}{\giga\hertz}$, an incoherent multi-frequency image, and a coherent multi-frequency image based on \eqref{eq:sumF} are shown in Fig.~\ref{fig:plane_res}(a), (b), and (c), respectively. The results are presented as 2-D maximum intensity projections (MIP) in the $xy$-plane, demonstrating the capability of the proposed ISR-based imaging method for modulated signals. The phase-corrected superposition provides significant improvements in terms of reduced artifacts, enhanced resolution, and high-contrast imaging.

\begin{figure}[t]
	\centering
	\subfloat[]{\includegraphics[scale=0.5]{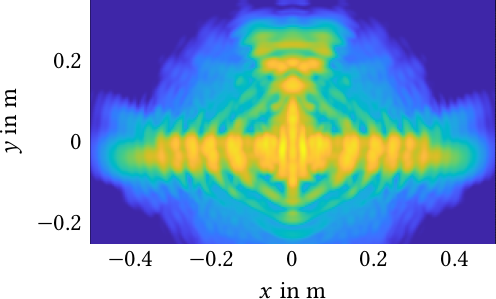}}%
	\hfill
	\subfloat[]{\includegraphics[scale=0.5]{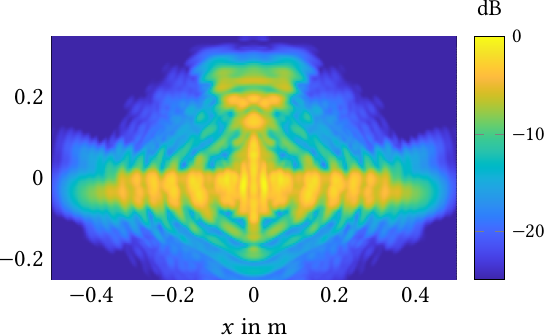}}%
	\caption{Imaging results of the PEC aircraft under narrowband signal illumination, (a)~incoherent multi-frequency image, and (b)~coherent multi-frequency image.}
	\label{fig:plane_nb}
\end{figure}

However, in realistic passive radar applications, such wideband signals may not be available. To investigate the influence of bandwidth, another simulation was performed with a narrow bandwidth ranging from $\SI{8}{\giga\hertz}$ to $\SI{8.1}{\giga\hertz}$ and a step size of $\SI{2.5}{\mega\hertz}$ to keep the total number of frequency samples the same. The imaging results for incoherent summation and coherent summation are shown in Fig.~\ref{fig:plane_nb}(a) and (b), respectively. It is seen that both imaging results are significantly worse than in the wideband case, and the effect of phase correction under the narrowband condition is also marginal. The limited spectral resources lead to degraded artifact suppression in the multi-frequency superposition and also result in a poor imaging resolution. An approximate expression for the imaging resolution can be found in~\cite{wang2024TAP,lopez-sanchez20003d}, while a more accurate treatment in the NF case is provided in~\cite{ahmed_electronic_2014}.


\section{Measurements}
\label{sec:meas}
\begin{figure}[t]
	\centering
	\includegraphics[scale=0.34]{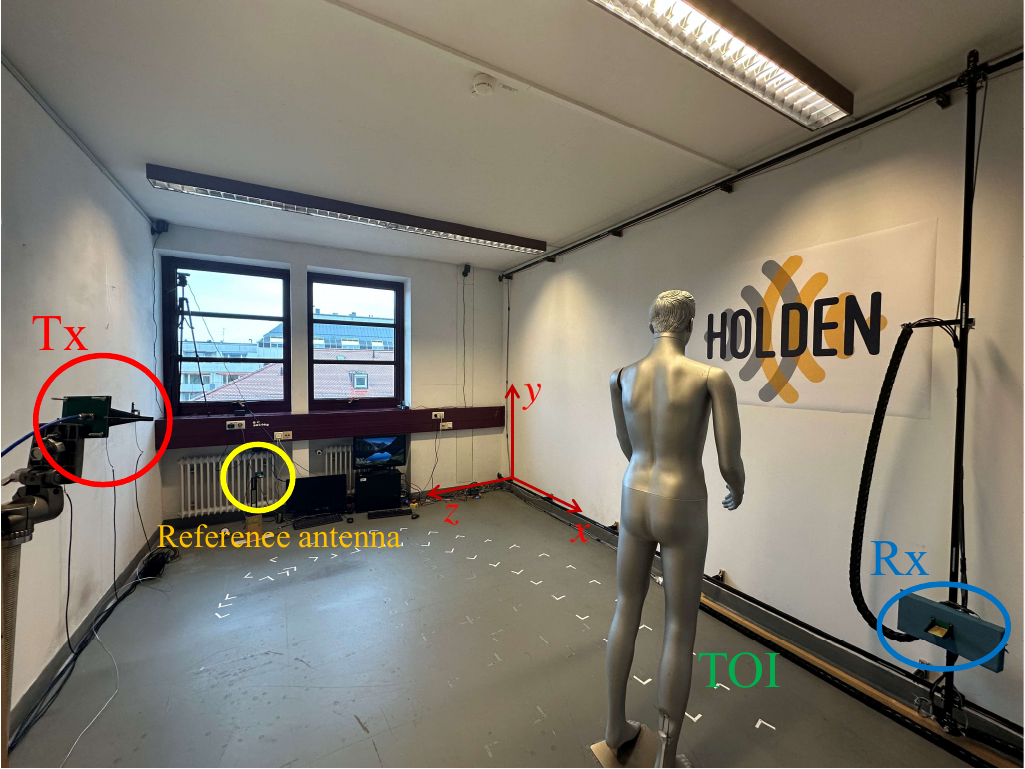}
	\caption{Measurement configuration utilized in a typical office room, consisting of a fixed illuminating Tx antenna, a fixed reference antenna, and an Rx antenna mounted on a positioner for planar measurements.}
	\label{fig:meas_env}
\end{figure}%
A measurement campaign was conducted in a typical office room at the Chair of High-Frequency Engineering, Technical University of Munich, as shown in Fig.~\ref{fig:meas_env}. For the Tx, an ADALM Pluto software defined radio (SDR)~\cite{Pluto2025} connected to a Vivaldi antenna was employed to transmit orthogonal frequency-division multiplexing (OFDM) signals centered around the carrier frequency at $\SI{2.41}{\giga \hertz}$ with a maximum bandwidth of $\SI{10}{\mega\hertz}$. For the Rx, an SDR based on an AD9361~\cite{AD9361} with two coherent receive channels and with a sampling rate of $\SI{15.36}{\mega \hertz}$ was employed. The fixed reference antenna and the moving probe antenna were connected to the SDR and automatic gain control of the receive channels was disabled. Sets of $\num{96e3}$ I/Q samples of both channels, corresponding to a measurement duration of $\SI{6.25}{\milli \second}$, were captured and converted to the frequency domain via a Fourier transform to obtain the harmonic magnitude and phase at each subcarrier frequency. The TOI was a mannequin coated with zinc-aluminum spray and placed at three different positions in the room. Owing to the highly reflective indoor environment, a background measurement without the TOI, while keeping all other setup conditions unchanged, was first performed for background subtraction~\cite{wang2024indoor}. The planar measurements were performed with a sampling step of approximately $\lambda/4$ which lead to $162\times 80$ sampling points uniformly distributed over a rectangular area of approximately $\SI{4.98}{\meter}\times \SI{2.46}{\meter}$ on the right side wall of the room. 
\begin{figure}[t]
	\centering
	\subfloat[]{\includegraphics[scale=0.47]{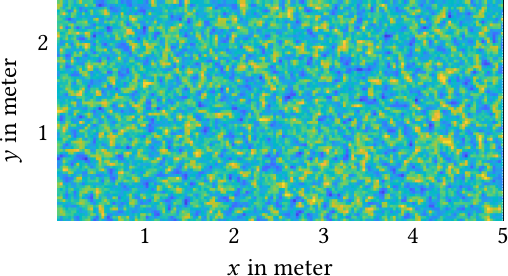}}%
	\hfill
	\subfloat[]{\includegraphics[scale=0.47]{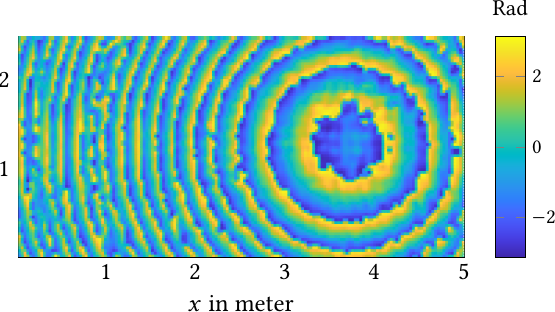}}%
  \caption{Phase distributions in radian at $\SI{2.41}{\giga \hertz}$ of the background measurement, (a)~without normalization and (b)~normalized by the reference signals.}
  \label{fig:meas_phase}
\end{figure}

\begin{figure}[t]
	\centering
	\subfloat[]{\includegraphics[scale=0.74]{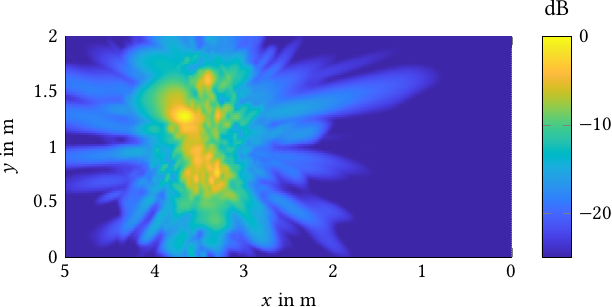}}%
	\hfill
  \\
  \vspace{-7mm}
	\subfloat[]{\includegraphics[scale=0.74]{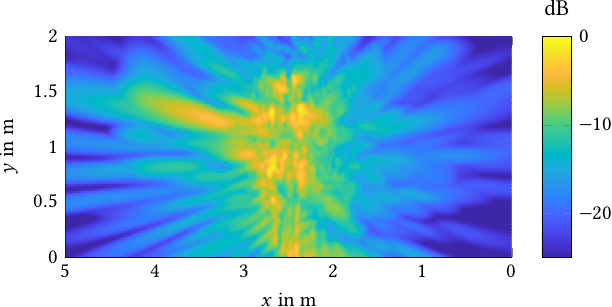}}%
  \hfill
  \\
  \vspace{-7mm}
	\subfloat[]{\includegraphics[scale=0.74]{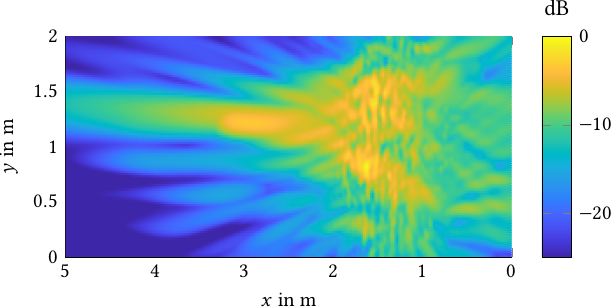}}%
  \caption{Single-frequency images of the mannequin at three different positions, with the Vivaldi antennas configured for vertical polarization.}
  \label{fig:man_pos}
\end{figure}

\begin{figure}[t]
	\centering
	\subfloat[]{\includegraphics[scale=0.74]{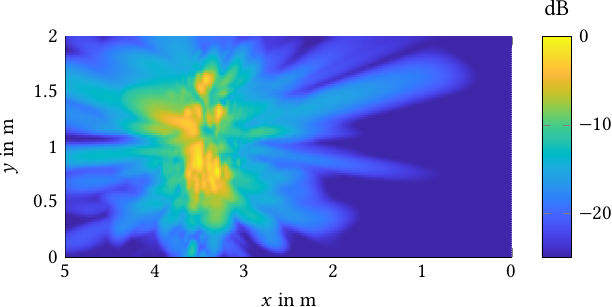}}%
	\hfill
  \\
  \vspace{-7mm}
	\subfloat[]{\includegraphics[scale=0.74]{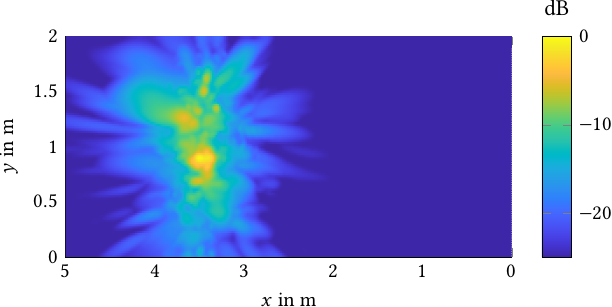}}%
    \caption{Coherent multi-frequency images of the mannequin obtained by combining 21 subcarriers within a $\SI{10}{\mega\hertz}$ bandwidth, with the Vivaldi antennas configured for (a) vertical polarization, and (b) horizontal polarization.}
  \label{fig:man_mulF}
\end{figure}

Illustrations of the NF measurement data at $\SI{2.41}{\giga \hertz}$ are given in Fig.\,\ref{fig:meas_phase}. The phase comparison between the raw measurement data and the results after normalization with the reference signals clearly shows that spatial coherence is lost due to the subsequent sampling over time and lack of Tx-Rx synchronization but can be recovered through normalization with the reference signals. First, the Vivaldi antennas of the Tx, the Rx, and the reference antenna were all set to vertical polarization, and single-frequency imaging at the center frequency is considered. By applying the ISR and image generation to the normalized measurements, the imaging results after background subtraction are obtained as shown in Fig.~\ref{fig:man_pos} for three different mannequin positions. Although strong artifacts are present, the human phantom can still be recognized in all three cases. The stronger artifacts observed when the mannequin is placed toward the right side are likely due to stronger scattering contributions from structures such as the windows and the heaters on the right side of the office room, as can be seen in Fig.~\ref{fig:meas_env}. In addition, 21 subcarriers around the center frequency within a $\SI{10}{\mega\hertz}$ bandwidth are coherently combined for multi-frequency imaging while incoherent superposition exhibits similar performance. A comparison between vertical and horizontal polarization is provided in Fig.~\ref{fig:man_mulF}. As observed in the results and supported by the simulations, the improvement from multi-frequency imaging is marginal due to the narrowband illumination. The horizontal polarization exhibits fewer artifacts, likely due to weaker interactions and interference in the testing environment. It is expected that clearer improvements in terms of the image quality may be obtained from data with small bandwidth, if the wave propagation within the considered echoic environment is more accurately modeled, e.g., in terms of a suitable numerical Green's function obtained from ray-tracing computations~\cite{Na.2023}.


%


\section{Conclusion}
\label{sec:conclusion}
An ISR based NF imaging method utilizing modulated signals for passive radar applications has been presented. By employing a fixed reference antenna, the NF measurements are normalized with respect to the reference signals, thereby enabling spatial coherence. The method is formulated in terms of wave propagation, where the effects of normalization and the corresponding phase correction are explicitly addressed. Simulations have verified the effectiveness of the phase correction. Although it was found that the benefit of the phase correction is relatively small for narrowband signals, there is strong potential for passive radar applications when sufficient spectral resources are available, e.g., $\SI{60}{\giga\hertz}$ Wi-Fi according to IEEE 802.11ad/ay. Furthermore, measurements in a typical office room using SDR-based Tx and Rx confirm that imaging with modulated signals via ISR can be successfully achieved with the aid of a reference antenna and background subtraction.


\section*{ACKNOWLEDGEMENT}
Funded by the European Union. Views and opinions expressed are however those of the author(s) only and do not necessarily reflect those of the European Union or European Innovation Council and SMEs Executive Agency (EISMEA). Neither the European Union nor the granting authority can be held responsible for them. Grant Agreement No: 101099491.



%

\bibliographystyle{IEEEtran}
\bibliography{Literature.bib}

@article{griffiths2010klein,
  title = {Klein {Heidelberg}{\textemdash}{The} First Modern Bistatic Radar System},
  author = {Griffiths, Hugh and Willis, Nicholas},
  year = {2010},
  month = oct,
  journal = {IEEE Trans. Aerosp. Electron. Syst.},
  volume = {46},
  number = {4},
  pages = {1571--1588},
  issn = {0018-9251},
  doi = {10.1109/TAES.2010.5595580},
  urldate = {2024-01-15}
}

@article{kuschel2019tutorial,
  title = {Tutorial: Passive Radar Tutorial},
  shorttitle = {Tutorial},
  author = {Kuschel, Heiner and Cristallini, Diego and Olsen, Karl Erik},
  year = {2019},
  month = feb,
  journal = {IEEE Aerosp. Electron. Syst. Mag.},
  volume = {34},
  number = {2},
  pages = {2--19},
  issn = {0885-8985, 1557-959X},
  doi = {10.1109/MAES.2018.160146},
  urldate = {2024-01-15}
}

@article{colone2023passive,
  title = {Passive Radar: Past, Present, and Future Challenges},
  shorttitle = {Passive Radar},
  author = {Colone, Fabiola and Filippini, Francesca and Pastina, Debora},
  year = {2023},
  month = jan,
  journal = {IEEE Aerosp. Electron. Syst. Mag.},
  volume = {38},
  number = {1},
  pages = {54--69},
  issn = {1557-959X},
  doi = {10.1109/MAES.2022.3221685},
  urldate = {2023-10-30}
}

@inproceedings{huang2014feasibility,
  title = {Feasibility and Limits of {Wi-Fi} Imaging},
  booktitle = {Proc. 12th ACM Conf. Embed. Netw. Sens. Syst.},
  author = {Huang, Donny and Nandakumar, Rajalakshmi and Gollakota, Shyamnath},
  year = {2014},
  month = nov,
  pages = {266--279},
  address = {New York, NY, USA},
  doi = {10.1145/2668332.2668344},
  urldate = {2023-09-14},
  isbn = {978-1-4503-3143-2}
}

@inproceedings{adib2013see,
  title = {See through Walls with {WiFi}!},
  booktitle = {Proc. ACM SIGCOMM Conf. SIGCOMM},
  author = {Adib, Fadel and Katabi, Dina},
  year = {2013},
  month = aug,
  pages = {75--86},
  address = {New York, NY, USA},
  doi = {10.1145/2486001.2486039},
  urldate = {2023-10-26},
  isbn = {978-1-4503-2056-6}
}

@article{huang2022passive,
  title = {Passive Multistatic Radar Imaging of Vessel Target Using {GNSS} Satellites of Opportunity},
  author = {Huang, Chuan and Li, Zhongyu and An, Hongyang and Sun, Zhichao and Wu, Junjie and Yang, Jianyu},
  year = {2022},
  journal = {IEEE Trans. Geosci. Remote Sens.},
  volume = {60},
  pages = {1--16},
  issn = {1558-0644},
  doi = {10.1109/TGRS.2022.3195993},
  urldate = {2023-10-27}
}

@article{samczynski20225g,
  title = {{5G} Network-Based Passive Radar},
  author = {Samczy{\'n}ski, Piotr and Abratkiewicz, Karol and P{\l}otka, Marek and Zieli{\'n}ski, Tomasz P. and Wszo{\l}ek, Jacek and Hausman, S{\l}awomir and Korbel, Piotr and Ksi{\c e}{\.z}yk, Adam},
  year = {2022},
  journal = {IEEE Trans. Geosci. Remote Sens.},
  volume = {60},
  pages = {1--9},
  issn = {1558-0644},
  doi = {10.1109/TGRS.2021.3137904},
  urldate = {2023-10-30}
}

@article{holl2017holography,
  title = {Holography of {Wi}-{Fi} Radiation},
  author = {Holl, Philipp M. and Reinhard, Friedemann},
  year = {2017},
  month = may,
  journal = {Phys. Rev. Lett.},
  volume = {118},
  number = {18},
  pages = {183901},
  publisher = {American Physical Society},
  doi = {10.1103/PhysRevLett.118.183901},
  urldate = {2023-07-06}
}

@article{vakalis2019imaging,
  title = {Imaging With {WiFi}},
  author = {Vakalis, Stavros and Gong, Liang and Nanzer, Jeffrey A.},
  year = {2019},
  journal = {IEEE Access},
  volume = {7},
  pages = {28616--28624},
  issn = {2169-3536},
  doi = {10.1109/ACCESS.2019.2902315},
  urldate = {2023-10-30}
}

@inproceedings{wang2025automatic,
  title = {An Automatic Focusing Technique for Inverse Source Reconstructions Based Near-Field Passive Radar Imaging},
  booktitle = {19th Eur. Conf. Antennas Propag. (EuCAP)},
  author = {Wang, Quanfeng and Eibert, Thomas F.},
  year = {2025},
  month = mar,
  address = {Stockholm, Sweden},
  pages = {1--5},
  doi = {10.23919/EuCAP63536.2025.10999544},
  urldate = {2025-05-22}
}

@inproceedings{wang2024microwavea,
  title = {Microwave Imaging of Electromagnetic Wave Absorbers in an Antenna Measurement Chamber},
  booktitle = {IEEE Int. Symp. Antennas Propag.},
  author = {Wang, Quanfeng and Eibert, Thomas F.},
  year = {2024},
  month = jul,
  pages = {1139--1140},
  issn = {1947-1491},
  address = {Florence, Italy},
  doi = {10.1109/AP-S/INC-USNC-URSI52054.2024.10686558},
  urldate = {2024-10-04}
}

@article{eibert2015electromagnetic,
  title = {Electromagnetic Field Transformations for Measurements and Simulations (Invited Paper)},
  author = {Eibert, Thomas F. and Kilic, Emre and Lopez, Carlos and Mauermayer, Raimund A. M. and Neitz, Ole and Schnattinger, Georg},
  year = {2015},
  journal = {Prog. Electromagn. Res.},
  volume = {151},
  pages = {127--150},
  issn = {1559-8985},
  doi = {10.2528/PIER14121105},
  urldate = {2023-12-13},
  langid = {english}
}

@article{schnattinger2012solution,
  title = {Solution to the Full Vectorial {3}-{D} Inverse Source Problem by Multilevel Fast Multipole Method Inspired Hierarchical Disaggregation},
  author = {Schnattinger, Georg and Eibert, Thomas F.},
  year = {2012},
  month = jul,
  journal = {IEEE Trans. Antennas Propag.},
  volume = {60},
  number = {7},
  pages = {3325--3335},
  issn = {1558-2221},
  doi = {10.1109/TAP.2012.2196946},
  urldate = {2023-11-13}
}

@article{eibert2023inverse,
  title = {Inverse Source Solutions with Simultaneous Localization in the Spatial and Spectral Domains{\textendash}Sparse Sampling for Directive Antennas},
  author = {Eibert, Thomas F. and Saurer, Matthias M. and Paulus, Alexander H. and Knapp, Josef},
  year = {2023},
  journal = {IEEE Trans. Antennas Propag.},
  pages = {779--790},
  volume = {1},
  month = jan,
  doi = {10.1109/TAP.2023.3322561},
  urldate = {2023-12-05}
}

@Book{Chew2001,
  title     = {Fast and Efficient Algorithms in Computational Electromagnetics},
  publisher = {Artech House},
  author    = {Weng C. Chew and Jian-Ming Jin and Eric Michielssen and Jiming Song},
  address   = {Boston},
  year      = {2001},
  language  = {english},
}

@Electronic{EMSS2024,
  author    = {Altair},
  title     = {{FEKO}},
  language  = {english},
  url       = {https://altairhyperworks.com/product/FEKO},
  year      = {2025},
  publisher = {Altair Engineering},
}

@Electronic{Pluto2025,
  author    = {{Analog Devices}},
  title     = {{ADALM-PLUTO}},
  language  = {english},
  url       = {https://wiki.analog.com/university/tools/pluto},
  year      = {2025},
}

@Electronic{AD9361,
  author    = {{Analog Devices}},
  title     = {{AD9361 RF Transceiver}},
  language  = {english},
  url       = {https://www.analog.com/en/products/ad9361.html},
  year      = {2013},
}

@article{lopez-sanchez20003d,
  title = {{3-D} Radar Imaging Using Range Migration Techniques},
  author = {{Lopez-Sanchez}, J.M. and {Fortuny-Guasch}, J.},
  year = {2000},
  month = may,
  journal = {IEEE Trans. Antennas Propag.},
  volume = {48},
  number = {5},
  pages = {728--737},
  issn = {1558-2221},
  doi = {10.1109/8.855491},
  urldate = {2024-05-06}
}

@inproceedings{Na.2023,
 author = {Na, Han and Saurer, Matthias and Eibert, Thomas F.},
 title = {Electromagnetic ray tracing simulation and imaging of complex indoor scenarios},
 pages = {300--304},  
 booktitle = {25th Int. Conf. Electromagn. Adv. Appl. (ICEAA)},
 year = {2023}, 
 address = {Venice, Italy},
 month = {October}
}

@book{ahmed_electronic_2014,
	title = {Electronic {Microwave} {Imaging} with {Planar} {Multistatic} {Arrays}},
	isbn = {978-3-8325-3621-3},
	language = {en},
	publisher = {Logos Verlag Berlin GmbH},
	author = {Ahmed, Sherif Sayed},
	year = {2014},
  address   = {Berlin},
}

@article{wang2024indoor,
  title = {An Indoor Localization Technique Utilizing Passive Tags and {3-D} Microwave Passive Radar Imaging},
  author = {Wang, Quanfeng and Paulus, Alexander H. and Tong, Mei Song and Eibert, Thomas F.},
  year = {2024},
  journal = {Prog. Electromagn. Res.},
  volume = {181},
  pages = {89--98},
  issn = {1559-8985},
  doi = {10.2528/PIER24120903},
  urldate = {2024-12-30},
  langid = {english}
}

@book{newton2002scattering,
  title = {Scattering Theory of Waves and Particles: Second Edition},
  shorttitle = {Scattering Theory of Waves and Particles},
  author = {Newton, Roger G.},
  year = {2002},
  month = nov,
  edition = {2nd edition},
  publisher = {Dover Publications Inc.},
  address = {Mineola, N.Y},
  isbn = {978-0-486-42535-1},
  langid = {english}
}

@article{wang2024TAP,
  title = {{3-D} Near-Field Passive Radar Imaging Utilizing Phase-Corrected Frequency Domain Inverse Source Reconstructions},
  author = {Wang, Quanfeng and Saurer, Matthias M. and Eibert, Thomas F.},
  year = {2025},
  month = jan,
  journal = {IEEE Trans. Antennas Propag.},
  volume = {73},
  number = {1},
  pages = {504--516},
  issn = {1558-2221},
  doi = {10.1109/TAP.2024.3498438},
  urldate = {2025-01-21}
}

\end{document}